\documentclass{article}
\usepackage{spconf,amsmath,graphicx,siunitx,bbm}
\usepackage{booktabs}
\usepackage{tikz}
\usepackage{pgfplots}
\usepgfplotslibrary{groupplots}
\pgfplotsset{compat=1.12}
\usepackage{multirow}
\usepackage{csvsimple}
\usepackage{hyperref}
\usepackage{float}
\usepackage{makecell}

\def\etal{et al.~}

\newcommand{\Fig}[1] {Fig.~\ref{#1}}

\definecolor{bblue}{HTML}{4F81BD}
\definecolor{rred}{HTML}{C0504D}
\definecolor{ggreen}{HTML}{9BBB59}
\definecolor{ppurple}{HTML}{9F4C7C}
\pgfkeys{/pgf/number format/.cd,1000 sep={}}

\title{DeepFilterNet: A Low Complexity Speech Enhancement Framework for Full-Band Audio based on Deep Filtering}
\name{Hendrik Schröter$^1$, Alberto N. Escalante-B.$^2$, Tobias Rosenkranz$^2$, Andreas Maier$^1$}
\address{
  $^1$Friedrich-Alexander-Universit\"at Erlangen-N\"urnberg, Pattern Recognition Lab\\
  $^2$WS Audiology, Research and Development, Erlangen, Germany
}
\begin{document}
%
\maketitle
\begin{abstract}
  Complex-valued processing has brought deep learning-based speech enhancement and signal extraction to a new level.
  Typically, the process is based on a time-frequency (TF) mask which is applied to a noisy spectrogram, while complex masks (CM) are usually preferred over real-valued masks due to their ability to modify the phase.
  Recent work proposed to use a complex filter instead of a point-wise multiplication with a mask.
  This allows to incorporate information from previous and future time steps exploiting local correlations within each frequency band.

  In this work, we propose DeepFilterNet, a two stage speech enhancement framework utilizing deep filtering.
  First, we enhance the spectral envelope using ERB-scaled gains modeling the human frequency perception.
  The second stage employs deep filtering to enhance the periodic components of speech.
  Additionally to taking advantage of perceptual properties of speech, we enforce network sparsity via separable convolutions and extensive grouping in linear and recurrent layers to design a low complexity architecture.

  We further show that our two stage deep filtering approach outperforms complex masks over a variety of frequency resolutions and latencies and demonstrate convincing performance compared to other state-of-the-art models.

\end{abstract}
\section{Introduction}
\label{sec:intro}

Monaural speech enhancement is an important part in many systems such as automatic speech recognition, video conference systems, as well as assistive listening devices.
Most state-of-the-art approaches work in the short-time Fourier transform (STFT) representation and estimate a TF mask using a deep neural network, many of these either real-valued masks \cite{valin2018rnnoise,valin2020perceptually,zhang21tlowdelay} or complex masks \cite{williamson2016complex,tan2019complex,roux2019phasebook,lv2021dccrnplus}.
The estimated masks are usually well-defined and limited by an upper bound to improve stability of the network training.
However, typically both approaches degrade if the frequency resolution gets to low for removing noise between speech harmonics.
The approaches above work on at least \SI{20}{\ms} windows resulting in a minimum frequency of \SI{50}{\Hz}.

In this paper, we propose an open source speech enhancement framework based on deep filtering (DF) \cite{mack2019deep, schroeter2020clcnet}.
Instead of using a complex mask that is applied per TF-bin, we use a combination of real-valued gains and a deep filter enhancement component.
For the first stage, we take advantage from the fact that noise as well as speech usually have a smooth spectral envelope.
An equivalent rectangular bandwidth (ERB) filter bank reduces input and output dimensions to only 32 bands, allowing for a computationally cheap encoder/decoder network.
Since the resulting minimum bandwidth of \SI{100}{\Hz} to \SI{250}{\Hz} depending on the FFT size is typically not sufficient to enhance periodic components, we use a second enhancement stage based on deep filtering.
That is, a deep filter network estimates coefficients for frequency bins up to an upper frequency $f_\text{DF}$.
The resulting linear complex-valued filters are applied to their corresponding frequency bins.
DF enhancement is only applied for lower frequencies since periodic speech components contain most energy in the lower frequencies.

Deep filtering was first proposed by Mack \etal\cite{mack2019deep} and Schröter \etal\cite{schroeter2020clcnet}.
Since a filter applied to multiple time/frequency (TF) bins, DF is able to recover signal degradations like notch-filters or time-frame zeroing.
Schröter \etal\cite{schroeter2020clcnet} introduced this method as complex linear coding (CLC) for low latency hearing aid applications.
CLC was motivated by its ability to model quasi-static properties of speech.
That is, even for frequency bandwidth of \SI{500}{\Hz} CLC is able to reduce noise within a frequency band, while preserving speech components.
This is especially helpful, when there are multiple speech harmonics within one frequency bin or for filtering periodic noises.
Recent work \cite{lv2021dccrnplus} demonstrated good performance using deep filtering in the deep noise suppression challenge \cite{reddy2021interspeech}.
However, compared to their previous work \cite{hu2020dccrn} using a complex ratio mask (CRM), their improvements are mostly given by network architecture changes like complex TF-LSTMs or convolutional pathways.

In this work, we demonstrate superior performance of deep filtering over CRMs for multiple FFT sizes from \SI{5}{\ms} to \SI{30}{\ms}.
We further show that even for low latency requirements of e.g.\ \SI{5}{\ms} resulting in a frequency resolution of \SI{250}{\Hz}, DF can still enhance the periodic speech components.

\section{{DeepFilterNet}}
\label{sec:deepfilternet}

\vspace{-.25em}\subsection{Signal Model}
\label{ssec:signalmodel}
Let $x(t)$ be a mixture signal recorded in a noisy room.
\vspace{-.25em}\begin{equation}
  x(t) = s(t) * h(t) + z(t)
\end{equation}
where $s(t)$ is a clean speech signal, $h(t)$ is a room impulse response from the speaker to the microphone and $z(t)$ is an additive noise signal already containing reverberation.
Typically, noise reduction operates in frequency domain:
\begin{equation}
  X(k, f) = S(k, f) \cdot H(k, f) + Z(k, f)\text{,}
\end{equation}
where $X(k, f)$ is the STFT representation of the time domain signal $x(t)$ and $k$, $f$ are the time and frequency bins.

\vspace{-.25em}\subsection{Deep Filtering}

Deep filtering is defined by a complex filter in TF-domain:
\vspace{-0.5em}
\begin{equation}
  \vspace{-0.5em}
  Y(k, f) = \sum_{i=0}^{N} C(k, i, f) \cdot X(k - i + l, f)\text{\ ,}
  \label{eq:DF}
\end{equation}
where $C$ are the complex coefficients of filter order $N$ that are applied to the input spectrogram $X$, and $\hat{Y}$ the enhanced spectrogram.
In our framework, the deep filter is applied to the gain-enhanced spectrogram $Y^G$.
$l$ is an optional look-ahead, which allows incorporating non-causal taps in the linear combination if $l\ge1$.
Additionally, one could also filter over the frequency axis allowing to incorporate correlations e.g.\ due to overlapping bands.
To further make sure that deep filtering only affects periodic parts, we introduce a learned weighting factor $\alpha$ to produce the final output spectrogram:
\vspace{-.25em}\begin{equation}
  Y^{DF}(k, f) = \alpha(k) \cdot Y^{DF'}(k, f) + (1-\alpha(k)) \cdot Y^G(k, f)\text{.}
  \vspace{-.25em}
\end{equation}

\vspace{-.5em}\subsection{Framework Overview}
\label{sec:framework}

\begin{figure}[tb]
  \vspace{-.25em}
  \centering
  \includegraphics[width=.67\linewidth, trim=3.5cm 5.2cm 10cm 0.2cm, clip]{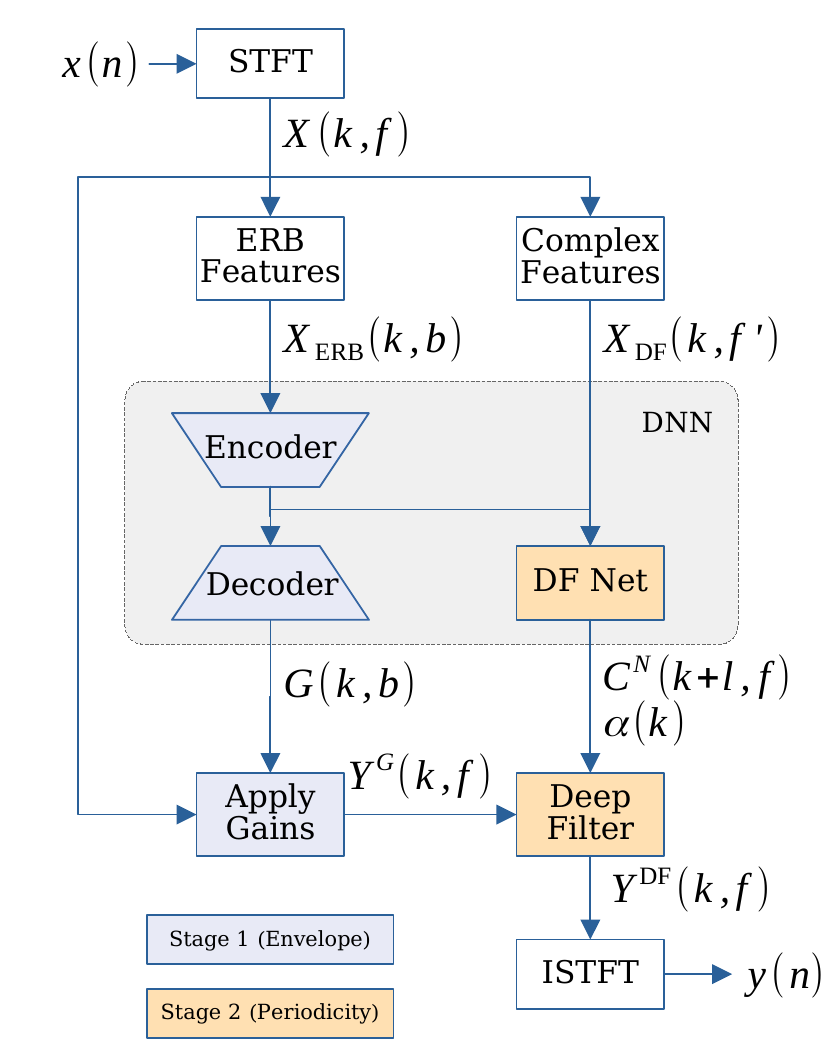}
  \vspace{-.5em}
  \caption{Overview of the DeepFilterNet algorithm. Stage 1 blocks are indicated in blue, stage 2 blocks in yellow.}
  \label{fig:deepfilternet}
  \vspace{-.5em}
\end{figure}

An overview of the DeepFilterNet algorithm is shown in \Fig{fig:deepfilternet}.
Given a noisy audio signal $x(t)$ we transform the signal into frequency domain using a short time Fourier transform (STFT).
The framework is designed for sampling rates up to \SI{48}{\kHz} to support high resolution VoIP applications and STFT window sizes $N_{FFT}$ between \SI{5}{\ms} and \SI{30}{\ms}.
By default, we use an overlap of $N_{ov}=\SI{50}{\percent}$ but also support higher overlaps.
We use two kinds of input features for the deep neural network (DNN).
For the ERB encoder/decoder features $X_{\text{ERB}}(k, b)$, $b\in[0,N_{\text{ERB}}]$, we compute a log-power spectrogram, normalize it using an exponential mean normalization \cite{schroeter2020hcrnn} with decay of \SI{1}{\second} and apply a rectangular ERB filter bank (FB) with a configurable number of bands $N_\text{ERB}$.
In fact, this normalization is similar to using instance normalization like \cite{lv2021dccrnplus}, which also estimates statistics with a momentum-based approach and only adds additional scaling and bias parameters.
For the deep filter network features $X_\text{DF}(k, f')$, $f'\in[0,f_\text{DF}]$, we use the complex spectrogram as input and normalize it using an exponential unit normalization \cite{schroeter2020clcnet} with the same decay.

An encoder/decoder architecture is used to predict ERB-scaled gains.
An inverse ERB filter bank is applied to transform the gains back to frequency domain before pointwise multiplying with the noisy spectrogram.
To further enhance the periodic components, DeepFilterNet predicts per-band filter coefficients $C^N$ of order $N$.
We only utilize Deep Filtering up to a frequency $f_\text{DF}$ assuming that the periodic components contain most energy in lower frequencies.

Together with the DNN look-ahead in the convolutional layers as well as the deep filter look-ahead, the overall latency is given by
$l_{N_\text{FFT}} + \max(l_\text{DNN},\ l_\text{DF})$ resulting in a minimal latency of $5+\max(0, 0)=\SI{5}{\ms}$ for $N_{\text{FFT}}=240$.

\vspace{-.2em}
\subsection{DNN Model}
\label{ssec:DNN}

\begin{figure}[bt]
  \includegraphics[width=\linewidth,trim=0cm 0cm 0cm 0cm,clip, page=2]{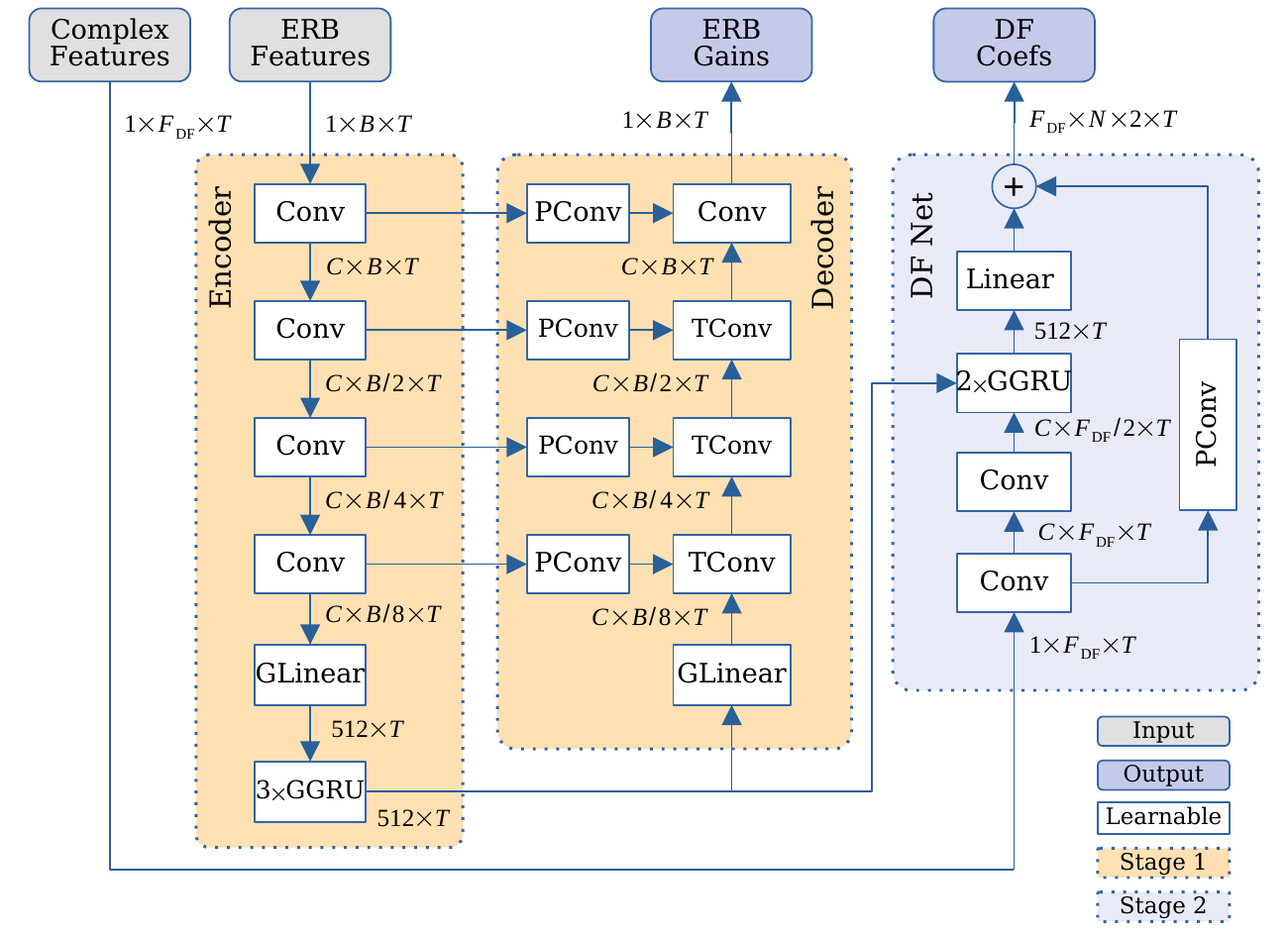}
  \vspace{-1.5em}
  \caption{
    Overview of the DeepFilterNet architecture.
    We use 1x1 pathway convolutions (PConv) as add-skip connections and transposed convolutional blocks (TConv) analogous to the encoder blocks.
    Grouped linear and GRU (GLinear, GGRU) layers are used to introduce sparsity.
  }
  \label{fig:DNN}
  \vspace{-1em}
\end{figure}

We focus on designing an efficient DNN only using standard DNN layers like convolutions, batch normalization, ReLU, etc., so that we can take advantage of layer fusing as well utilize good support by inference frameworks.
We adopt a UNet-like architecture similar to \cite{braun2021towards, lv2021dccrnplus} as shown in \Fig{fig:DNN}.
Our convolutional blocks consist of a separable convolutions (depthwise followed by a 1x1 convolution) with kernel size of (3x2) and $C=64$ channels followed by a batch normalization and ReLU activation.
The convolutional layers are aligned in time such that the first layers may introduce an overall look-ahaead $l_\text{DNN}$.
The remaining convolutional layers are causal and do not contribute any more latency.
We heavily make use of grouping \cite{tan2019learning, braun2021towards} for our linear and GRU layers.
That is, the layer input is split into $P=8$ groups resulting in $P$ smaller GRUs/linear layers with a hidden size of $512/P=64$.
The output is shuffled to recover inter-group correlations and concatenated to the full hidden size again.
Convolutional pathways \cite{braun2021towards, lv2021dccrnplus} with add-skips are used to retain frequency resolution.
We use a global pathway skip connection for the DF Net to provide a good representation of the original noisy phase at the output layer.

\subsection{Data Preprocessing}

The DeepFilterNet framework utilizes heavy on-the-fly augmentation.
We mix a clean speech signal with up to 5 noise signals at signal-to-noise ratios (SNR) of $\{-5, 0, 5, 10, 20, 40\}$ \si{\dB}.
To further increase variabilty, we augment speech as well as noise signals with second order filters \cite{valin2018rnnoise}, EQs, and random gains of $\{-6, 0, 6\}$ \si{\dB}.
Random resampling increases the variety of pitches and room impulse responses (RIR) are used for simulating reverberant environments.
We apply a low-pass filter to noise signals before mixing, if the sampling rate of a speech signal is lower than the current model's sampling rate.
This e.g.\ also allows models trained on full-band audio (\SI{48}{\kHz}) to perform equally well on input signals with lower sampling rates.
We furthermore support training attenuation limited models.
Therefore, we generate a ``noisy'' target signal $s$ with a \num{6} to \num{20} \si{\dB} higher SNR compared to the noisy signal $x$.
During training, we then clamp the predicted gains $G$ and having a ``noisy'' target $s$, DF Net will learn to not remove more noise than specified.
This is useful e.g.\ for wearable devices where we want to keep some environmental awareness for the user.

\subsection{Loss}
\label{ssec:loss}

It is not trivial to provide ideal DF coefficients $C^N$, since there are infinitely many possibilities \cite{mack2019deep}.
Instead, we use a compressed spectral loss to implicitly learn ERB gains $G$ and filter coefficients $C^N$ \cite{ephrat2018looking, braun2021towards}.
\begin{equation}
  \mathcal{L}_{spec} =
  \sum_{k, f} ||\ |Y|^c - |S|^c ||^2 + \sum_{k, f} ||\ |Y|^c e^{j\varphi_Y} - |S|^c e^{j\varphi_S} ||^2\text{,}
\end{equation}
where $c=0.6$ is a compression factor to model the perceived loudness \cite{valin2021echocontrol}.
Having a magnitude as well as a phase-aware term makes this loss is suitable for modeling both real-valued gain and complex DF coefficient prediction.
To harden the gradient for TF bins with magnitude close to zero (e.g.\ for input signals with a lower sampling rate), we compute the angle backward method of $\varphi_X$ like:
\vspace{-.25em}\begin{equation}
  \frac{\delta\varphi}{\delta X} = \delta X\cdot(\frac{-\Im\{X\}}{|X_h|^2}, \frac{\Re\{X\}}{|X_h|^2})\text{,}
\end{equation}
where $\Re\{X\}$ and $\Im\{X\}$ represent real and imaginary part of spectrogram $X$ and $|X_h|^2=\max(\Re\{X\}^2 + \Im\{X\}^2, 1e^{-12})$ is the hardened squared magnitude to avoid by \num{0} division.

\begin{figure*}[t]
  \vspace{-.25em}
  \centering
  \hspace{-0.2cm}\begin{tikzpicture}
    \begin{groupplot}[
      ybar,
      legend entries={DF,CRM},
      legend to name=CombinedLegendBar,
      footnotesize,
      ybar legend,
      width=.33\linewidth,
      height=4.0cm,
      grid=both,
      grid style={solid,gray!30!white},
      every axis plot/.append style={fill},
      every axis/.append style={
          bar width=7pt 
        },
      xlabel={FFT size},
      ylabel={SI-SDR [\si{\dB}]},
      y tick label style={
          /pgf/number format/.cd,%
          scaled y ticks = false,
          precision=3,
          set decimal separator={.},
          fixed},%
      xtick={240, 480, 720, 960, 1200, 1440},
      group style={
          group size=3 by 1,
          ylabels at=edge left,
        }]
      \nextgroupplot[title={\scriptsize SNR 0}]
      \addplot[ybar, draw=black,fill=bblue] table [x=fftsize, y=sisdr, col sep=comma] {assets/df_fft_snr0.csv};
      \addplot[ybar, draw=black,fill=rred] table [x=fftsize, y=sisdr, col sep=comma] {assets/crm_fft_snr0.csv};
      \nextgroupplot[title={\scriptsize SNR 5}]
      \addplot[ybar, draw=black,fill=bblue] table [x=fftsize, y=sisdr, col sep=comma] {assets/df_fft_snr5.csv};
      \addplot[ybar, draw=black,fill=rred] table [x=fftsize, y=sisdr, col sep=comma] {assets/crm_fft_snr5.csv};
      \nextgroupplot[title={\scriptsize SNR 10}]
      \addplot[ybar, draw=black,fill=bblue] table [x=fftsize, y=sisdr, col sep=comma] {assets/df_fft_snr10.csv};
      \addplot[ybar, draw=black,fill=rred] table [x=fftsize, y=sisdr, col sep=comma] {assets/crm_fft_snr10.csv};
    \end{groupplot}
    \node[right=0.1cm, inner sep=0pt] at(current bounding box.east) {\pgfplotslegendfromname{CombinedLegendBar}};
  \end{tikzpicture}
  \vspace{-1.0em}
  \caption{Comparison of Deep Filtering (DF) and conventional complex ratio masks (CRM) over multiple FFT sizes corresponding to \num{5} to \SI{30}{\ms}.}
  \label{fig:comparison_fft}
  \vspace{-1em}
\end{figure*}
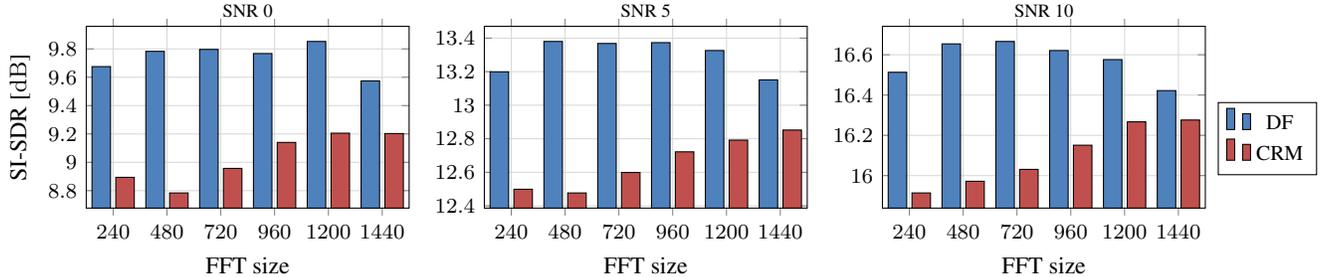

As an additional loss term, we force the DF component to only enhance periodic parts of the signal.
The motivation is as follows.
DF does not provide any benefit over ERB gains for noise-only sections.
DF may even cause artifacts by modeling periodic noises like engine or babble noise which is most noticeable in attenuation limited models.
Furthermore, DF does not provide any benefit for speech with only stochastic components like fricatives or plosives.
Assuming, that these sections contain most of the energy in higher frequencies, we compute the local SNR for frequencies below $f_\text{DF}$.
Therefore, $\mathcal{L}_\alpha$ is given by
\vspace{-.25em}\begin{equation}
  \mathcal{L}_\alpha = \sum_k ||\alpha \cdot \mathbbm{1}_{\text{LSNR}<\SI{-10}{\dB}}||^2 + \sum_k ||(1-\alpha) \cdot \mathbbm{1}_{\text{LSNR}>\SI{-5}{\dB}}||^2\text{,}
\end{equation}
where $\mathbbm{1}_{\text{LSNR}<\SI{-10}{\dB}}$ is the characteristic function with value \num{1} if the local SNR (LSNR) is smaller than \SI{-10}{\dB} and $\mathbbm{1}_{\text{LSNR}>\SI{-5}{\dB}}$ \num{1} if LSNR is greater than \SI{-5}{\dB}.
The LSNR is computed in STFT domain up to a frequency of $f_\text{DF}$ over \SI{20}{\ms} windows.
This ensures that DF is only applied in segments containing significant amount of speech energy in low frequencies.
The combined loss is given by
\begin{equation}
  \mathcal{L} = \lambda_{spec}\cdot\mathcal{L}_{spec}(Y, S) + \lambda_{\alpha}\cdot \mathcal{L}_\alpha\text{.}
\end{equation}

\section{Experiments}
\vspace{-.2em}\subsection{Training setup}

We train our models based on the deep noise suppression (DNS) challenge dataset \cite{reddy2021interspeech} containing over \SI{750}{\hour} of full-band clean speech and \SI{180}{\hour} of various noise types.
We oversample included high-quality speech datasets, VCTK and PTDB, by a factor of 10.
Additionally to the provided RIRs which are sampled at \SI{16}{\kHz}, we simulate another \num{10000} RIRs at \SI{48}{\kHz} using the image source model \cite{habets2007generating} with RT60s of \num{0.05} to \SI{1.00}{\second}.
The VCTK and PTDB datasets are split on speaker level ensuring no overlap with the VCTK test set \cite{valentini2016investigating}, the remaining dataset is split at signal level into train/validation/test (\num{70}/\num{15}/\SI{15}{\percent}).
Early stopping is applied based on the validation loss, results are reported on the test set.
The VCTK/DEMAND test set \cite{valentini2016investigating} is used to compare DeepFilterNet to related work.

\begingroup
\setlength{\thinmuskip}{1mu}
\setlength{\medmuskip}{2mu plus 1mu minus 4mu}
\setlength{\thickmuskip}{2mu plus 2mu}
All experiments use full-band signals with a sampling rate of \SI{48}{\kHz}.
We take $N_\text{ERB}=32$, $f_\text{DF}=\SI{5}{kHz}$, DF order of $N=5$, and a look-ahead of $l_\text{DF}=1$ and $l_\text{DNN}=2$ for the convolutions.
We train our models on \SI{3}{\second} samples and a batch size of \num{32} for \num{30} epochs using an Adam optimizer with an initial learning rate of \num{1e-3}.
The learning rate is decayed by a factor of \num{0.9} every \num{3} epochs.
Loss parameters are $\lambda_{spec}=1$ and $\lambda_\alpha=0.05$.
The framework source code can be obtained at \url{https://github.com/Rikorose/DeepFilterNet}.
\endgroup

\vspace{-.2em}
\subsection{Results}
We evaluate our framework over multiple FFT sizes and compare the performance of DF and CRMs based on the scale-invariant signal-distortion-ratio (SI-SDR) \cite{le2019sdr}.
CRM is just a special case of DF, where order $N=1$ and look-ahead $l=0$.
The DNN look-ahead remains the same for the CRM models.

\begin{figure}[b]
  \centering
  \vspace{-1.25em}
  \includegraphics[width=\linewidth]{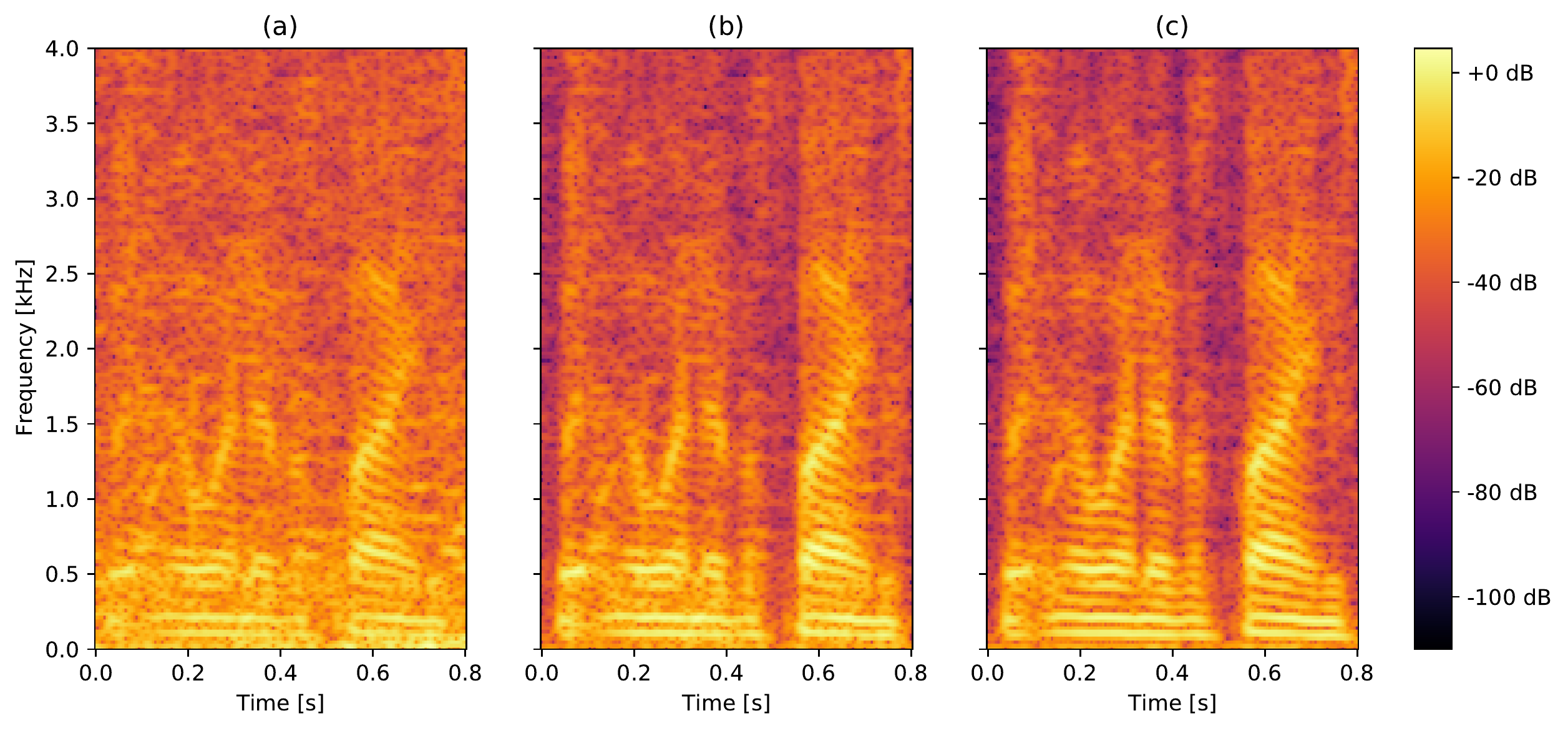}
  \vspace{-1.4em}
  \caption{Sample from the VCTK test set. Noisy (a), CRM (b), DF (c) ($N_\text{FFT}=960$ for CRM and DF).}
  \label{fig:zoom}
\end{figure}
\Fig{fig:comparison_fft} shows that DF outperforms CRM over all FFT sizes corresponding to \SI{5}{\ms} to \SI{30}{\ms}.
Limited by the frequency resolution, the performance of CRMs drops for FFT window sizes $\le\SI{20}{\ms}$.
On the other hand, the relatively constant performance of DF drops around \SI{30}{\ms} due to a smaller amount of correlation in neighboring frames.
Increasing the FFT overlap to \SI{75}{\percent} results in a slightly better performance for both, DF and CRM ($+\SI{0.6}{\dB}$ SI-SNR for input SNR 0).
This performance increase can be explained by a higher intra-frame correlation as well as the DNN having twice as many steps to update the RNN hidden states at the cost of doubling the computational complexity.
\Fig{fig:zoom} shows a qualitative example to demonstrate the capabilities of DF to reconstruct speech harmonics that are indistinguishable in the noisy spectrogram.
\begin{table}[bt]
  \vspace{-.65em}
  \caption{Objective results on VCTK/DEMAND test set. Unreported values of related work are indicated as ``-''.}
  \label{tab:voicebank_test}
  \centering
  \robustify\bfseries
  \sisetup{
    table-number-alignment = center,
    table-figures-integer  = 1,
    table-figures-decimal  = 2,
    table-auto-round = true,
    detect-weight = true
  }
  \resizebox{\linewidth}{!} {
    \begin{tabular}{
        l S[table-figures-decimal=1] S S S
      }%
      \toprule%
      Model                                  & \multicolumn{1}{c}{\makecell{Params                                                        \\ $[\text{M}]$}} & \multicolumn{1}{c}{\makecell{MACS \\ $[\text{G}]$}} & \multicolumn{1}{c}{\makecell{WB-PESQ\\ $[\text{MOS}]$}} & \multicolumn{1}{c}{\makecell{SI-SDR \\ $[\text{dB}]$}}\\%
      \midrule%
      Noisy                                  & \multicolumn{1}{c}{-}               & \multicolumn{1}{c}{-} & 1.97 & 8.41                  \\%
      PercepNet \cite{valin2020perceptually} & 8.0                                 & 0.8                   & 2.73 & \multicolumn{1}{c}{-} \\%
      DCCRN \cite{hu2020dccrn}               & 3.7                                 & 14.36                 & 2.68 & \multicolumn{1}{c}{-} \\%
      DCCRN+ \cite{lv2021dccrnplus}          & 3.3                                 & \multicolumn{1}{c}{-} & 2.84 & \multicolumn{1}{c}{-} \\%
      DeepFilterNet                          & 1.778                               & 0.3482                & 2.81 & 16.63                 \\%
      \hspace{.3em} w/o stage 2              & 0.885                               & 0.2506                & 2.57 & 13.81123              \\
      \bottomrule%
    \end{tabular}
  }
  \vspace{-1em}
\end{table}

We compare DeepFilterNet with $N_\text{FFT}=960$ (\SI{20}{\ms}) to related work like PercepNet \cite{valin2020perceptually} which uses a similar perceptual approach as well as DCCRN+ \cite{lv2021dccrnplus} which also utilizes deep filtering.
We assess speech enhancement quality using WB-PESQ \cite{ITU2007WBPESQ} and compare computational complexity in multiply-and-accumulate per second (MACS).
Table \ref{tab:voicebank_test} shows that DeepFilterNet outperforms PercepNet and performs on par with DCCRN+ while having a much lower computational complexity making DeepFilterNet viable for real-time usage.


\section{Conclusion}
\label{sec:conclusion}

In this work, we proposed DeepFilterNet, a low complexity speech enhancement framework.
We showed that DeepFilterNet performs on par with other algorithms, while being computationally more efficient.
This is achieved using a perceptually motivated approach allowing to minimize the model complexity.
Moreover, we provided evidence that DF outperforms CRMs, particularly for smaller STFT window sizes.
In the future, we plan to further improve the perceptual approach by better applying DF to periodic components of speech e.g.~using a correlation-based voiced probability.

\bibliographystyle{IEEEbib}
\bibliography{refs.bib}

\begin{thebibliography}{10}

\bibitem{valin2018rnnoise}
Jean-Marc Valin,
\newblock ``{A hybrid DSP/deep learning approach to real-time full-band speech
  enhancement},''
\newblock in {\em 2018 IEEE 20th International Workshop on Multimedia Signal
  Processing (MMSP)}. IEEE, 2018, pp. 1--5.

\bibitem{valin2020perceptually}
Jean-Marc Valin, Umut Isik, Neerad Phansalkar, Ritwik Giri, Karim Helwani, and
  Arvindh Krishnaswamy,
\newblock ``{A Perceptually-Motivated Approach for Low-Complexity, Real-Time
  Enhancement of Fullband Speech},''
\newblock in {\em INTERSPEECH 2020}, 2020.

\bibitem{zhang21tlowdelay}
Xu~Zhang, Xinlei Ren, Xiguang Zheng, Lianwu Chen, Chen Zhang, Liang Guo, and
  Bing Yu,
\newblock ``{Low-Delay Speech Enhancement Using Perceptually Motivated Target
  and Loss},''
\newblock in {\em Proc. Interspeech 2021}, 2021.

\bibitem{williamson2016complex}
Donald~S Williamson, Yuxuan Wang, and DeLiang Wang,
\newblock ``Complex ratio masking for monaural speech separation,''
\newblock {\em IEEE/ACM Transactions on Audio, Speech and Language Processing
  (TASLP)}, vol. 24, no. 3, pp. 483--492, 2016.

\bibitem{tan2019complex}
Ke~Tan and DeLiang Wang,
\newblock ``Complex spectral mapping with a convolutional recurrent network for
  monaural speech enhancement,''
\newblock in {\em ICASSP 2019-2019 IEEE International Conference on Acoustics,
  Speech and Signal Processing (ICASSP)}. IEEE, 2019, pp. 6865--6869.

\bibitem{roux2019phasebook}
Jonathan Le~Roux, Gordon Wichern, Shinji Watanabe, Andy Sarroff, and John~R
  Hershey,
\newblock ``Phasebook and friends: Leveraging discrete representations for
  source separation,''
\newblock {\em IEEE Journal of Selected Topics in Signal Processing}, vol. 13,
  no. 2, pp. 370--382, 2019.

\bibitem{lv2021dccrnplus}
Shubo Lv, Yanxin Hu, Shimin Zhang, and Lei Xie,
\newblock ``{DCCRN+: Channel-wise Subband DCCRN with SNR Estimation for Speech
  Enhancement},''
\newblock in {\em INTERSPEECH}, 2021.

\bibitem{mack2019deep}
Wolfgang Mack and Emanu{\"e}l~AP Habets,
\newblock ``{Deep Filtering: Signal Extraction and Reconstruction Using Complex
  Time-Frequency Filters},''
\newblock {\em IEEE Signal Processing Letters}, vol. 27, pp. 61--65, 2020.

\bibitem{schroeter2020clcnet}
Hendrik Schröter, Tobias Rosenkranz, Alberto Escalante~Banuelos, Marc
  Aubreville, and Andreas Maier,
\newblock ``{CLCNet}: {Deep} learning-based noise reduction for hearing aids
  using complex linear coding,''
\newblock in {\em ICASSP 2020-2020 IEEE International Conference on Acoustics,
  Speech and Signal Processing (ICASSP)}, 2020.

\bibitem{reddy2021interspeech}
Chandan~KA Reddy, Harishchandra Dubey, Kazuhito Koishida, Arun Nair, Vishak
  Gopal, Ross Cutler, Sebastian Braun, Hannes Gamper, Robert Aichner, and
  Sriram Srinivasan,
\newblock ``{INTERSPEECH 2021 Deep Noise Suppression Challenge},''
\newblock in {\em INTERSPEECH}, 2021.

\bibitem{hu2020dccrn}
Yanxin Hu, Yun Liu, Shubo Lv, Mengtao Xing, Shimin Zhang, Yihui Fu, Jian Wu,
  Bihong Zhang, and Lei Xie,
\newblock ``{DCCRN: Deep complex convolution recurrent network for phase-aware
  speech enhancement},''
\newblock in {\em INTERSPEECH}, 2020.

\bibitem{schroeter2020hcrnn}
Hendrik Schröter, Tobias Rosenkranz, Alberto~N. Escalante-B., Pascal Zobel,
  and Andreas Maier,
\newblock ``{Lightweight} {Online} {Noise} {Reduction} on {Embedded} {Devices}
  using {Hierarchical} {Recurrent} {Neural} {Networks},''
\newblock in {\em INTERSPEECH 2020}, 2020.

\bibitem{braun2021towards}
Sebastian Braun, Hannes Gamper, Chandan~KA Reddy, and Ivan Tashev,
\newblock ``Towards efficient models for real-time deep noise suppression,''
\newblock in {\em ICASSP 2021-2021 IEEE International Conference on Acoustics,
  Speech and Signal Processing (ICASSP)}. IEEE, 2021, pp. 656--660.

\bibitem{tan2019learning}
Ke~Tan and DeLiang Wang,
\newblock ``Learning complex spectral mapping with gated convolutional
  recurrent networks for monaural speech enhancement,''
\newblock {\em IEEE/ACM Transactions on Audio, Speech, and Language
  Processing}, vol. 28, pp. 380--390, 2019.

\bibitem{ephrat2018looking}
Ariel Ephrat, Inbar Mosseri, Oran Lang, Tali Dekel, Kevin Wilson, Avinatan
  Hassidim, William~T Freeman, and Michael Rubinstein,
\newblock ``{Looking to Listen at the Cocktail Party: A Speaker-Independent
  Audio-Visual Model for Speech Separation},''
\newblock {\em ACM Transactions on Graphics (TOG)}, vol. 37, no. 4, pp. 1--11,
  2018.

\bibitem{valin2021echocontrol}
Jean-Marc Valin, Srikanth Tenneti, Karim Helwani, Umut Isik, and Arvindh
  Krishnaswamy,
\newblock ``{Low-Complexity, Real-Time Joint Neural Echo Control and Speech
  Enhancement Based On PercepNet},''
\newblock in {\em {2021 IEEE International Conference on Acoustics, Speech and
  Signal Processing (ICASSP)}}. IEEE, 2021.

\bibitem{habets2007generating}
Emanu{\"e}l~AP Habets and Sharon Gannot,
\newblock ``Generating sensor signals in isotropic noise fields,''
\newblock {\em The Journal of the Acoustical Society of America}, vol. 122, no.
  6, pp. 3464--3470, 2007.

\bibitem{valentini2016investigating}
Cassia Valentini-Botinhao, Xin Wang, Shinji Takaki, and Junichi Yamagishi,
\newblock ``{Investigating RNN-based speech enhancement methods for
  noise-robust Text-to-Speech},''
\newblock in {\em SSW}, 2016, pp. 146--152.

\bibitem{le2019sdr}
Jonathan Le~Roux, Scott Wisdom, Hakan Erdogan, and John~R Hershey,
\newblock ``{SDR--half-baked or well done?},''
\newblock in {\em ICASSP 2019-2019 IEEE International Conference on Acoustics,
  Speech and Signal Processing (ICASSP)}. IEEE, 2019, pp. 626--630.

\bibitem{ITU2007WBPESQ}
{ITU},
\newblock ``{Wideband extension to Recommendation P.862 for the assessment of
  wideband telephone networks and speech codecs},''
\newblock {\em ITU-T Recommendation P.862.2}, 2007.

\end{thebibliography}

\end{document}